\documentclass[12pt]{article}
\usepackage{epsfig}
\usepackage{epstopdf}

\usepackage{amssymb}
\usepackage{amsmath}
\usepackage{amsfonts}
\usepackage{amsthm}
\usepackage{graphicx}
\usepackage{mathrsfs}
\usepackage{multirow}

\newcommand{\brho}{\boldsymbol{\rho}}
\newcommand{\bsigma}{\boldsymbol{\sigma}}

\newcommand{\R}{\mathbb{R}}
\newcommand{\C}{\mathbb{C}}

\newcommand{\fm}{\mathfrak{m}}
\newcommand{\fn}{{\mathfrak{n}}}

\newcommand{\fz}{\mathfrak{z}}

\newcommand{\fK}{\mathfrak{K}}

\newcommand{\bk}{\mathbf{k}}

\newcommand{\bH}{\mathbf{H}}
\newcommand{\bI}{\mathbf{I}}

\newcommand{\bM}{\mathbf{M}}

\newcommand{\bS}{\mathbf{S}}

\newcommand{\cH}{\mathcal{H}}

\newcommand{\cP}{\mathcal{P}}

\newcommand{\cT}{\mathcal{T}}
\newcommand{\cU}{\mathcal{U}}

\newcommand{\be}{\begin{equation}}
\newcommand{\ee}{\end{equation}}
\newcommand{\bea}{\begin{eqnarray}}
\newcommand{\eea}{\end{eqnarray}}
\newcommand{\nn}{\nonumber}
\newcommand{\kt}{\rangle}
\newcommand{\br}{\langle}

\newcommand{\ed}{\end{document}}

\newcommand{\bi}{\begin{itemize}}
\newcommand{\ei}{\end{itemize}}

\newcommand{\bce}{\begin{center}}
\newcommand{\ece}{\end{center}}

\newcommand{\sH}{\mathscr{H}}

\newcommand{\sP}{\mathscr{P}}
\newcommand{\sR}{\mathscr{R}}

\newcommand{\bcH}{{\boldsymbol{\cH}}}
\newcommand{\bcU}{{\boldsymbol{\cU}}}

\newcommand{\for}{{\mbox{\rm for}}}

\begin{document}
 
\title{Non-Stationary Quantum Systems Associated with Stationary Scattering in One Dimension}

\author{Ali~Mostafazadeh$^{1,2,}$\thanks{Corresponding author, E-mail address:
amostafazadeh@ku.edu.tr}~, and Alireza Shariati Joni$^{2,}$\thanks{E-mail address: ajoni25@ku.edu.tr} \\[6pt]
Departments of Mathematics$^{1}$ and Physics$^{2}$, Ko\c{c} University,\\  34450 Sar{\i}yer, Istanbul, T\"urkiye}

\date{ }
\maketitle

\begin{abstract}
Stationary scattering admits a dynamical formulation in which the scattering data are extracted from the evolution operator for an effective non-stationary and non-unitary quantum system. This provides a mathematical method of solving scattering problems. We consider the possibility of realizing such a quantum system and relating the scattering information to outcomes of quantum measurements performed on it. To this end, we examine stationary scattering problems in one dimension where the scattering data consist of reflection and transmission amplitudes and the corresponding effective quantum system is a non-unitary two-level system. The states of this system identify points in a Bloch sphere $S^2$ and their evolution yields trajectories in $S^2$. Quantum measurements turn out to be capable of providing information about the reflection amplitudes. We identify configurations displaying reflectionlessness with situations where one or both of the poles of $S^2$ undergo cyclic evolutions. Spectral singularities and coherent perfect absorption correspond to dynamical spin flips which map one of the poles of $S^2$ to the other.
\end{abstract}

\section{Introduction}
\label{s1}

Stationary scattering serves as the central framework for studying wave scattering phenomena. The standard approach to solve stationary scattering problems makes use of the Lippmann-Schwinger equation \cite{Lippmann-Schwinger,Newton,Sakurai}. In one dimension (1D), there is an alternative method that employs the concept of transfer matrix \cite{jones-1941,abeles,thompson}. This is a $2\times 2$ complex matrix whose entries determine the reflection and transmission amplitude of the problem. Transfer matrix methods provide a powerful tool for dealing with the scattering problems. Their central ingredient is the composition property of the transfer matrix which allows for slicing the scattering media into thin slices and assembling the contributions of the slices to determine the scattering features of the medium \cite{sanchez,tjp-2020}. This approach has been successfully utilized to address various effectively one-dimensional stationary scattering problems \cite{berreman-1972,pereyra-1998} and generalized to wave propagation and scattering in two and three dimensions \cite{pendry-1984,pendry-1990a,pendry-1990b}. 

The use of transfer matrices in the study of wave scattering in two and three dimensions involves identifying a scattering axis, slicing the scatterers along this axis, discretizing the transverse degrees of freedom for each slice, associating numerical transfer matrices with the slices, and determining the overall transfer matrix of the system by multiplying those of its individual slices in a certain order \cite{pendry-1996,McLean-Pendry}. The size of these transfer matrices depends on the data points used in the discretization scheme. While this approach provides a useful numerical framework for solving scattering problems, its main limitation is the multiplication of a large number of high-dimensional matrices, which can introduce significant numerical instability \cite{Luque-Raigon}.

There is a more recent approach to stationary scattering in two and three dimensions that involves certain integral operator generalization of the numerical transfer matrices and allows for analytic treatments of scattering problems \cite{pra-2021,pra-2023a,ptep-2026a}. The basic idea of this approach---which we call the dynamical formulation of stationary scattering (DFSS)--- is to map the stationary scattering phenomenon to the dynamics of an effective quantum system which is generally non-stationary and non-unitary. 

The introduction of DFSS was motivated by the observation that the composition rule for the classical transfer matrices of potential scattering in 1D coincides with that of the evolution operators in quantum mechanics. This led to the identification of a non-stationary two-level quantum system whose evolution operator determined the transfer matrix of the potential \cite{ap-2014,pra-2014a}. This observation has found various applications \cite{pra-2014a,jpa-2014a,pra-2014b,jmp-2021,jpa-2021} and generalizations \cite{epjp-2019,jpa-2020,ptep-2024b}. 

The purpose of the present article is to explore consequences of treating the effective quantum systems arising in DFSS in 1D as genuine quantum systems obeying von Neumann's measurement and projection postulates. In particular, we wish to explore the possibility of extracting information about the reflection and transmission amplitudes of a given scattering system by performing quantum measurements on the associated effective two-level quantum system.

The outline of the remainder of this article is as follows. In Sec.~2, we review DFSS in 1D potential scattering and its generalization to stationary scattering problems defined by Bergmann's equation in 1D, \cite{Bergmann}. The latter models the scattering of acoustic waves by inhomogeneities of a compressible fluid with planar symmetry \cite{Martin} as well as the scattering of transverse electric (TE) and transverse magnetic (TM) waves by scattering media with such a symmetry \cite{ptep-2024b}. In Sec.~3, we examine the dynamical and kinematical properties of the effective quantum systems appearing in the DFSS considered in Sec.~2 and explore the possibility of acquiring information about the reflection and transmission amplitudes by performing quantum measurements on these quantum systems. In Sec.~4, we examine the application of our general results for a few concrete toy models. In Sec.~5, we present  our concluding remarks.

\section{Dynamical formulation of stationary scattering in 1D}

Consider the stationary scattering problem defined by the time-independent Schr\"odinger equation,
	\be
	-\partial_x^2\psi(x)+v(x)\psi(x)=k^2\psi(x),
	\label{sch-eq}
	\ee
where $v:\R\to\C$ is a possibly complex-valued and energy-dependent short-range potential\footnote{This means that $v$ is absolutely integrable and satisfies $\lim_{x\to\pm\infty}xv(x)=0$.} and $k\in\R^+$ is the incident wavenumber. Then, for each solution $\psi$ of \eqref{sch-eq}, there are possibly $k$-dependent complex coefficients $A_\pm$ and $B_\pm$ such that
	\be
	\psi(x)\to A_\pm e^{ikx}+B_\pm e^{-ikx}~~~\for~~~x\to\pm\infty.
	\label{asym}
	\ee
Since $\psi$ defines a time-harmonic solution $e^{-i\omega t}\psi(x)$ of the corresponding time-dependent Schr\"odinger equation, with $\omega:=k^2$, $A_\pm$ and $B_\pm$ are the coefficients of the right-going and left-going waves at spatial infinities $x=\pm\infty$. 

The transfer matrix of the potential $v$ is the unique $k$-dependent $2\times 2$ matrix $\bM$ that satisfies
	\be
	\bM\left[\begin{array}{c}
	A_-\\
	B_-\end{array}\right]=\left[\begin{array}{c}
	A_+\\
	B_+\end{array}\right],
	\label{M-def}
	\ee
and is independent of $A_\pm$ and $B_\pm$, \cite{epjp-2019}. One can use the properties of \eqref{sch-eq} to prove that $\det \bM=1$, \cite{tjp-2020}. 

Let $\psi^{l/r}$ denote the left/right-incident scattering solutions of \eqref{sch-eq} which satisfy
	\begin{align}
	&\psi^l(x)\to N^{l}\times\left\{\begin{array}{ccc}
	e^{ikx}+R^l\, e^{-ikx}&\for&x\to-\infty,\\
	T^l\,e^{ikx}&\for&x\to\infty,\end{array}\right.
	\label{left}\\
	&\psi^r(x)\to N^{r}\times\left\{\begin{array}{ccc}
	T^r e^{-ikx}&\for&x\to-\infty,\\
	e^{-ikx}+R^r\, e^{ikx}&\for&x\to\infty,\end{array}\right.
	\label{right}
	\end{align}
where $N^{l/r},R^{l/r}$, and $T^{l/r}$ are respectively the amplitude of the incident wave, and left/right reflection and transmission amplitudes. Comparing \eqref{left} and \eqref{right} with \eqref{asym}, we can express the coefficients $A_\pm$ and $B_\pm$ associated with the left/right-incident waves in terms of $N^{l/r},R^{l/r}$, and $T^{l/r}$. Substituting these in \eqref{M-def} and making use of the fact that $\det\bM=1$, we find
	\begin{align}
	&R^l=-\frac{M_{21}}{M_{22}},
	&&R^r=\frac{M_{12}}{M_{22}},
	&&T^l=T^r=T:=\frac{1}{M_{22}}.
	\label{Rs-Ts}
	\end{align}
According to these equations $\bM$ stores the information about the reflection and transmission amplitudes. Therefore its determination is equivalent to the solution of the scattering problem for the potential $v$. 

Ref.~\cite{ap-2014} proves the following theorem which provides the conceptual basis for DFSS in 1D.
	\begin{itemize}
	\item[]{\bf Theorem~1:} Let $v:\R\to\C$ be a short-range potential, and $\bcH(x)$ be the following matrix-valued function of $x$.
		\be
		\bcH(x):=\frac{v(x)}{2k}\left[\begin{array}{cc}
		1 & e^{-2ikx}\\
		-e^{2ikx} & -1\end{array}\right].
		\label{H-def}
		\ee
Then the transfer $\bM$ associated with the potential $v$ is given by
	\be
	\bM=\bcU(\infty,-\infty):=\lim_{x_\pm\to\pm\infty}\bcU(x_+,x_-),
	\label{M=U}
	\ee
where for $x,x_0\in\R$, $\bcU(x,x_0)$ is the unique $2\times 2$ matrix-valued function of $x$ and $x_0$ that solves the  initial-value problem
	\begin{align}
	&i\partial_x \bcU(x,x_0)=\bH(x)\bcU(x,x_0),
	&&\bcU(x_0,x_0)=\bI,
	\label{U-def}
	\end{align}
and $\bI$ is the $2\times 2$ identity matrix.			
	\end{itemize}
If we view $x$ as an evolution (``time'') parameter, we can identify $\bcU(x,x_0)$ with the evolution operator for the two-level quantum system defined by the Hamiltonian $\bcH(x)$ in the Hilbert space $\sH:=\C^2$. It is easy to see that $\bcH(x)$ is not Hermitian. It actually fails to be diagonalizable. Furthermore, since it is a function of $x$, which we interpret as an effective ``time,'' the corresponding quantum system is non-stationary. This implies that, in general, $\bcU(x,x_0)$ and consequently $\bM$ do not admit closed-form expressions; we can only expand them in their  Dyson series. In particular,  
	\be
	\bM=\bI+\sum_{n=1}^\infty (-i)^n\int_{-\infty}^{\infty} dx_n\int_{-\infty}^{x_n} dx_{n-1}\cdots
	\int_{-\infty}^{x_2} dx_1\bcH(x_n)\bcH(x_{n-1})\cdots\bcH(x_1).
	\label{Dyson-1}
	\ee
	
If $v$ is a finite-range potential, i.e., there is an interval $[a_-,a_+]$ of real numbers such that $v(x)=0$ for $x\notin[a_-,a_+]$, $\bcH(x)$ vanishes for $x\notin[a_-,a_+]$, and \eqref{Dyson-1} becomes
	\be
	\bM=\bI+\sum_{n=1}^\infty (-i)^n\int_{a_-}^{a_+} dx_n\int_{a_-}^{x_n} dx_{n-1}\cdots
	\int_{a_-}^{x_2} dx_1\bcH(x_n)\bcH(x_{n-1})\cdots\bcH(x_1).
	\label{Dyson}
	\ee
Refs.~\cite{pra-2014a,jpa-2014a,pra-2014b,jmp-2021,jpa-2021} provide different applications of this formula.

Next, consider the Bergmann equation \cite{Bergmann} in 1D
	\begin{align}
	\partial_x[\alpha(x)^{-1}\partial_x\psi(x)]+\fK^2\beta(x)\psi(x)=0,
	\label{Bergmann-eq}
	\end{align}
where $\fK:=k|\cos\theta|$, $k$ is the incident wavenumber, $\theta$ is the angle between the incident wave vector $\bk_0$ and the positive $x$ axis, and $\alpha:\R\to\C$ and $\beta:\R\to\C$ are a pair of functions subject to the conditions \cite{ptep-2024b}:
	\begin{itemize}
	\item[]C1) $\alpha-1$ and $\beta-1$ are absolutely-integrable functions.
	\item[]C2) There is some $a\in\R^+$ such that $\alpha$ and $\beta$ are twice-differentiable
	outside the interval $[-a,a]$.
	\item[]C3) $\displaystyle \lim_{x\to\pm\infty}x[\alpha(x)-1]=\lim_{x\to\pm\infty}x[\beta(x)-1]=0$.\item[]C4) $\displaystyle\lim_{x\to\pm\infty}x\partial_x^j\alpha(x)=\lim_{x\to\pm\infty}x\partial_x^j\beta(x)=0$ for $j\in\{1,2\}$.
	\end{itemize}
Then, every solution of \eqref{Bergmann-eq} has the asymptotic form \eqref{asym}. In particular, \eqref{Bergmann-eq} admits left/right-incident scattering solutions, $\psi^{l}$ and $\psi^{r}$, satisfying \eqref{left} and \eqref{right}, respectively. These model the scattering of TE and TM waves by inhomogeneities of an effectively one-dimensional dielectric material \cite{ptep-2024b} and the scattering of acoustic waves in an effectively one-dimensional compressible fluid \cite{Norris}.\footnote{For the scattering of TE waves, $\alpha=\hat\mu$ and $\beta=\hat\varepsilon$, where $\hat\mu$ and $\hat\varepsilon$ are respectively the relative permeability and relative permittivity of the material. For the scattering of TM waves, $\alpha=\hat\epsilon$ and $\beta=\hat\mu$. For the scattering of acoustic waves, $\theta=0$, $\alpha$ is the mass density of the fluid, and $\beta=c_0^2\kappa$, where $c_0$ is the speed of sound at spatial infinity, and $\kappa$ is the compressibility of the fluid.} It is not difficult to see that we can define a corresponding transfer matrix using \eqref{M-def}, show that it has unit determinant, and express the reflection and transmission amplitudes in terms of its entries via \eqref{Rs-Ts}, \cite{ptep-2024b}.
 
 Note that for $\alpha(x)=1$, $\beta(x)=1-v(x)/\fK^2$, and $\theta=0$, which implies $\fK=k$, the Bergmann equation \eqref{Bergmann-eq} reduces to the Schr\"odinger equation \eqref{sch-eq}. Therefore the former is a generalization of the latter.

Ref.~\cite{ptep-2024b} proves the following generalization of Theorem~1 for the stationary scattering problems defined by \eqref{Bergmann-eq}.
	\begin{itemize}
	\item[]{\bf Theorem~2:} Let $\alpha:\R\to\C$ and $\beta:\R\to\C$ be functions fulfilling conditions C1-- C4, and $\bcH(x)$ be the following matrix-valued function of $x$.
	\begin{align}
	&\bcH(x):=\fK\left[\begin{array}{cc}
	-\fm_+(x)+1 
	&-\fm_-(x)\,e^{-2i\fK\, x}\\[6pt]
	\fm_-(x)\,e^{2i\fK\, x}&\fm_+(x)-1
	\end{array}\right],
	\label{H-def-2}
	\end{align}
where
	\begin{align}
	&\fm_\pm(x):=\frac{\sec^2\theta\,[\alpha(x)\beta(x)-1]\pm\alpha(x)^2+1}{2\alpha(x)}.
	\label{mmp=}
	\end{align}
Then the transfer matrix $\bM$ associated with the scattering problem defined by \eqref{Bergmann-eq} satisfies \eqref{M=U}, where $\bcU(x,x_0)$ is the unique $2\times 2$ matrix-valued function of $x$ and $x_0$ that solves \eqref{U-def}, i.e., $\bM$ is given by \eqref{Dyson-1}, and \eqref{Dyson} holds whenever $\alpha(x)=\beta(x)=1$ for $x\notin[a_-,a_+]$. 
	\end{itemize}
	
Theorems~1 and 2 provide a straightforward proof of the identity $\det\bM=1$. Since according to
\eqref{H-def} and \eqref{H-def-2}  $\bcH(x)$ is traceless, it follows that $\det\bcU(x,x_0)=1$, which together with \eqref{M=U} yields $\det\bM=1$.

\section{Scattering data from quantum measurements}

Since the Hamiltonians \eqref{H-def} and \eqref{H-def-2} are manifestly non-Hermitian, the corresponding effective quantum systems undergo non-unitary evolutions. This causes no problem in implementing von Neumann's measurement postulate \cite{von-Neumann} provided that we use the general expression for the Born rule that is invariant under arbitrary scalings, $\Psi\to c\Psi$ by nonzero complex numbers $c$, of a state vector $\Psi$ of the system that determines its state at the time of a measurement \cite{paper186}. 

Suppose that at ``time'' $x$ we wish to measure an observable $O$.\footnote{Observables are necessarily represented by Hermitian operators acting in the Hilbert space of the system \cite{paper186,Schechter}.}  According to von Neumann's projection postulate, the measurement can only yield one of the eigenvalues $\omega_\pm$ of $O$ with probability
	\be
	{\rm Pr}_{\omega_\pm}:=\frac{|\br\Phi_\pm|\Psi(x)\kt|^2}{\br\Psi(x)|\Psi(x)\kt},
	\label{Born}
	\ee
where $\Phi_-$ and  $\Phi_+$ are respectively eigenvectors of $O$ associated with the eigenvalues $\omega_-$ and $\omega_+$, which form an orthonormal basis of the Hilbert space, and $\Psi(t)$ is a state vector determining the state $\sR(x)$ of the quantum system immediately before the measurement \cite{paper186,Fock}.\footnote{Here we assume that $O$ is not a multiple of the identity operator, so that $\omega_-\neq\omega_+$.} We recall that the states of the system are one-dimensional subspaces of the Hilbert space $\sH:=\C^2$. We can identify these with the points of the projective Hilbert space $\sP(\sH):=\C P^1$ which has the structure of a two-dimensional sphere $S^2$. This is often called the Bloch sphere. In particular, the evolving state,
	\[\sR(x):=\big\{c\Psi(x)|c\in\C\big\},\]
defines a trajectory in $S^2$. 
	
In the following, we wish to explore the possibility of acquiring information about the reflection and transmission amplitudes of the scattering problems defined by the  Schr\"odinger and Bergmann equations, \eqref{sch-eq} and \eqref{Bergmann-eq} by performing measurements on the corresponding quantum systems. We confine our attention to finite-range scattering interactions, i.e., assume the existence of an interval $[a_-,a_+]$ of real numbers such that
	\begin{align}
	v(x)=0~~{\rm and}~~\alpha(x)=\beta(x)=1~~~\for~~~x\notin[a_-,a_+].
	\label{finite-range}
	\end{align}
Furthermore, following common practice, we identify the state vectors $\Psi$ and observables $O$ of the corresponding effective quantum systems with their matrix representation in the standard basis of $\C^2$, i.e., $\{(1,0),(0,1)\}$. In other words, we represent $\Psi$ and $O$ respectively by $2\times 1$ and $2\times 2$ complex matrices. 

We begin our analysis, by noting that according to \eqref{left} and \eqref{right}, the coefficients $A_\pm$ and $B_\pm$ appearing in \eqref{asym} that correspond to the left/right-incident scattering solutions $\psi^{l/r}$ have the following form.
	\begin{align}
	\mbox{For}~\psi^l\!:&~
	\begin{aligned}
	&A_-=N^l, &&B_-=N^lR^l, &&A_+=N^lT, &&B_+=0,
	\end{aligned}
	\label{left-coeef}\\
	\mbox{For}~\psi^r\!:&~
	\begin{aligned}
	&A_-=0, &&B_-=N^rT &&A_+=N^rR^r, &&B_+=N^r,
	\end{aligned}
	\label{right-coeef}
	\end{align}
where we have made use of the fact that $T^{l/r}=T$. Substituting \eqref{left-coeef} and \eqref{right-coeef} in \eqref{M-def} and noting that $N^{l/r}\neq 0$, we find
	\begin{align}
	&\bM\left[\begin{array}{cc}
	1\\
	R^l\end{array}\right]=\left[\begin{array}{cc}
	T\\
	0\end{array}\right],
	&&
	\bM\left[\begin{array}{cc}
	0\\
	T\end{array}\right]=\left[\begin{array}{cc}
	R^r\\
	1\end{array}\right].\nn
	\end{align}
It is easy to show that these relations are equivalent to
	\begin{align}
	&\bM\Phi_+=\left(T-\frac{R^lR^r}{T}\right)\Phi_+-\frac{R^l}{T}\,\Phi_-,
	&&\bM\Phi_-=\frac{R^r}{T}\,\Phi_++\frac{1}{T}\,\Phi_-,
	\label{M-LR}
	\end{align}
where 
	\begin{align}
	&\Phi_+:=\left[\begin{array}{cc}
	1\\
	0\end{array}\right],
	&&\Phi_-:=\left[\begin{array}{cc}
	0\\
	1\end{array}\right].
	\label{Phi-pm}
	\end{align}
	
Now, consider the evolution of the states of the effective quantum systems defined by the Hamiltonian operators \eqref{H-def} and \eqref{H-def-2} through the time-dependent Schr\"odinger equation, 
	\be
	i\partial_x\Psi(x)=\bcH(x)\Psi(x).
	\label{sch-eq-t}
	\ee
Setting $x_0=a_-$ and $\Psi(a_-)=\Phi_\pm$, we have 
	\be
	\Psi(x):=\Psi_\pm(x):=\bcU(x,a_-)\Phi_\pm.
	\label{Psi-pm}
	\ee
In view of \eqref{M=U}, this implies
	\be
	\Psi(a_+):=\Psi_\pm(a_+)=\bM\Phi_\pm.
	\label{Psi-pm-p}
	\ee
	
We can identify $\Phi_\pm$ with the eigenvectors of the observable defined by the Pauli matrix $\bsigma_3$ with eigenvalue $\omega_\pm:=\pm1$. Suppose that at ``time'' $x_0=a_-$ the system is in one of the eigenstates, 
	\be
	\sH_\pm:=\{c\,\Phi_\pm\,|\,c\in\C\,\},\nn
	\ee
of $\bsigma_3$, and that we wish to respectively compute the probabilities ${\rm Pr}_{1}^\pm$ and ${\rm Pr}_{-1}^\pm$ of finding the values $1$ and $-1$ if we measure $\bsigma_3$ at ``time'' $x=a_+$. 

According to \eqref{Psi-pm}, the state of the system at ``time'' $x$ is given by
	\be
	\sR_\pm(x):=\big\{ c\,\Psi_\pm(x)\,|\,c\in\C\,\}.
	\ee
We can use the fact that $\Psi_\pm(a_-)=\Phi_\pm\in\sH_\pm=\sR_\pm(a_-)$ together with Eqs.~\eqref{Born}, \eqref{M-LR}, and \eqref{Psi-pm-p} to show that
	\begin{align}
	&{\rm Pr}_{1}^+=\frac{|T^2-R^lR^r|^2}{|T^2-R^lR^r|^2+|R^l|^2},
	&&{\rm Pr}_{-1}^+=\frac{|R^l|^2}{|T^2-R^lR^r|^2+|R^l|^2},
	\label{prob-p}\\
	&{\rm Pr}_{1}^-=\frac{|R^r|^2}{1+|R^r|^{2}},
	&&{\rm Pr}_{-1}^-=\frac{1}{1+|R^r|^2},
	\label{prob-m}
	\end{align}
where we have made use of the fact that  $T\neq 0$, \cite{tjp-2020}.	Notice that, in light of \eqref{prob-m}, 
	\begin{align}
	&0\leq {\rm Pr}_{1}^-<1,
	&&0<{\rm Pr}_{-1}^-\leq 1.
	\end{align}
The limiting situation ${\rm Pr}_{1}^--1={\rm Pr}_{-1}^-=0$ is attained at the singularities of $R^r$ that lie on the positive real axis in the complex $k$ plane. These signify the emergence of spectral singularities \cite{prl-2009} which mark the onset of lasing in optical applications \cite{pra-2011a} and SASER (sound amplification by stimulated emission of radiation) in acoustics \cite{kend-2006}. 

In view of \eqref{prob-p} and \eqref{prob-m},  the left (respectively  right) reflectionlessness of the scattering interaction is associated with situations where ${\rm Pr}^+_{-1}=0$ (respectively ${\rm Pr}_{1}^-=0$). This means that $\sH_+$ (respectively $\sH_-$) evolves into $\sH_+$ (respectively $\sH_-$), i.e., $\sH_+$ (respectively $\sH_-$) is a cyclic state of the effective quantum system. According to \eqref{M-LR}, the corresponding state vector $\Phi_+$ (respectively $\Phi_-$) evolves into $T\,\Phi_+$ (respectively $T^{-1}\Phi_-$), i.e., it acquires a complex phase factor $T$ (respectively $T^{-1}$). If the scattering interaction is bidirectionally invisible, i.e., $R^{l/r}=0$ and $T=1$, then the state vectors $\Phi_\pm$ evolve into $\Phi_\pm$. In this case, all the states of the system undergo cyclic evolutions as $x$ changes from $a_-$ to $a_+$.

Next, we note that according to \eqref{prob-m},
	\be
	|R^r|^2=\frac{{\rm Pr}_{1}^-}{1-{\rm Pr}_{1}^-}.
	\label{eq1}
	\ee
Therefore, measurements of $\bsigma_3$ performed on the effective quantum system at time $x=a_+$ may be used to determine the right reflection coefficient $|R^r|^2$.\footnote{If $v$, $\alpha$, and $\beta$ are real-valued functions of $x$, $|R^l|=|R^r|$ and $|T|^2=1-|R^r|^2$, \cite{springer-2018}. Therefore these measurements determine both reflection coefficients $|R^{l/r}|^2$ and the transmission coefficient $|T|^2$.} This observation suggests that performing the same measurements on the effective quantum system obtained by performing the reflection $x\to \tilde x:=-x$ should allow for gaining information about $|R^l|^2$. 

To see this, first we recall that the reflected effective quantum systems are obtained by replacing $v(x)$, $\alpha(x)$, and $\beta(x)$ in \eqref{H-def} and \eqref{H-def-2} respectively with
	\begin{align}
	&\widetilde v(x):=v(\widetilde x)=v(-x),
	&&\widetilde \alpha(x):=\alpha(\widetilde x)=\alpha(-x),
	&&\widetilde \beta(x):=\beta(\widetilde x)=\beta(-x).\nn
	\end{align}
These vanish for $x\notin[\tilde a_-,\tilde a_+]$, where $\tilde a_\pm=-a_\mp$. As shown in Ref.~\cite{springer-2018}, the reflection amplitudes $\widetilde {R}^{l/r}$ of the reflected systems are related to those of the original systems via $\widetilde{R}^{l/r}= R^{r/l}$. Therefore, we can determine the left reflection coefficient $|R^l|^2$ using the probability ${\rm Pr}_{1}^-$ computed for the reflected quantum system with $a_\pm$ changed to $\tilde a_\pm=- a_\mp$. Denoting the latter quantity by $\widetilde{\rm Pr}_{\pm 1}^{_-}$ and using \eqref{eq1}, we have 
	\be
	|R^l|^2=|\widetilde R^r|^2=\frac{\widetilde{\rm Pr}_{1}^{_-}}{1-\widetilde{\rm Pr}_{1}^{_-}}.
	\label{eq2}
	\ee
	
An additional piece of information that we can extract from Eqs.~\eqref{prob-p}, \eqref{eq1}, and \eqref{eq2}, is the value of $|T^2-R^lR^r|^2$. The condition that $T^2-R^lR^r=0$ is equivalent to ${\rm Pr}^+_{-1}=1$ which means that the eigenstate $\sH_+$ evolves into its orthogonal complement $\sH_-$. This is usually called a spin flip. In the corresponding scattering setup, $T^2-R^lR^r=0$ is equivalent to $M_{11}=0$ which signifies the phenomenon of coherent perfect absorption \cite{tjp-2020,springer-2018,wan-2011,chong-2011}. In other words, coherent perfect absorption has the interpretation of a spin flip in the associated effective quantum system.

\section{Application to concrete toy models}
 
In this section, we examine the dynamics generated by the effective quantum systems defined by the Hamiltonian operators  \eqref{H-def} and \eqref{H-def-2} for some concrete choices for the functions $v$, $\alpha$, and $\beta$. More specifically, we provide graphical representations of the trajectories of the evolving states in the projective Hilbert space $\sP(\sH)=\C P^1=S^2$. To do this, we identify the states $\sR$ of the system with the normalized (unit-trace) density matrices, 
	\begin{equation}
	\brho=\frac{\Psi\,\Psi^\dagger}{\Psi^\dagger\Psi}=\frac{1}{2}(\bI+\vec r\cdot\vec\bsigma),
	\label{rho=}
	\end{equation}
where $\vec r=(r_1,r_2,r_3)\in\R^3$, $\vec\bsigma:=(\bsigma_1,\bsigma_2,\bsigma_3)$, and $\bsigma_j$ with $j\in\{1,2,3\}$ are Pauli matrices. Since, according to \eqref{rho=}, $\brho^2$ has unit trace, $|\vec r\,|^{2}=1$. Therefore, we can parameterize the states $\sR$ using the coordinates of the points on the unit sphere $S^2$ centered at the origin in $\R^3$. In particular, the states $\sH_+$ and $\sH_-$, which are respectively determined by the state vectors $\Phi_+$ and $\Phi_-$, correspond to the north (N) and south (S) poles of $S^2$.

\subsection{Scattering by a barrier potential with a Gaussian bump}

\begin{figure}
        {\begin{center}
        \includegraphics[scale=.6]{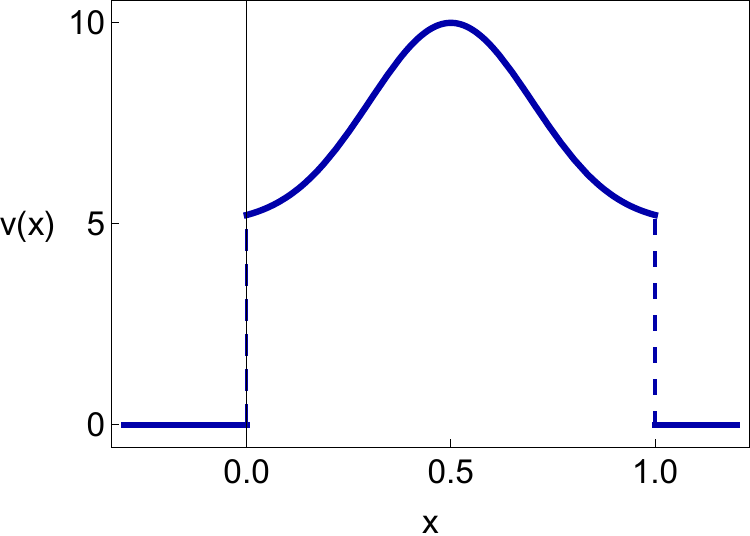}~~~~~~~~
        \includegraphics[scale=.45]{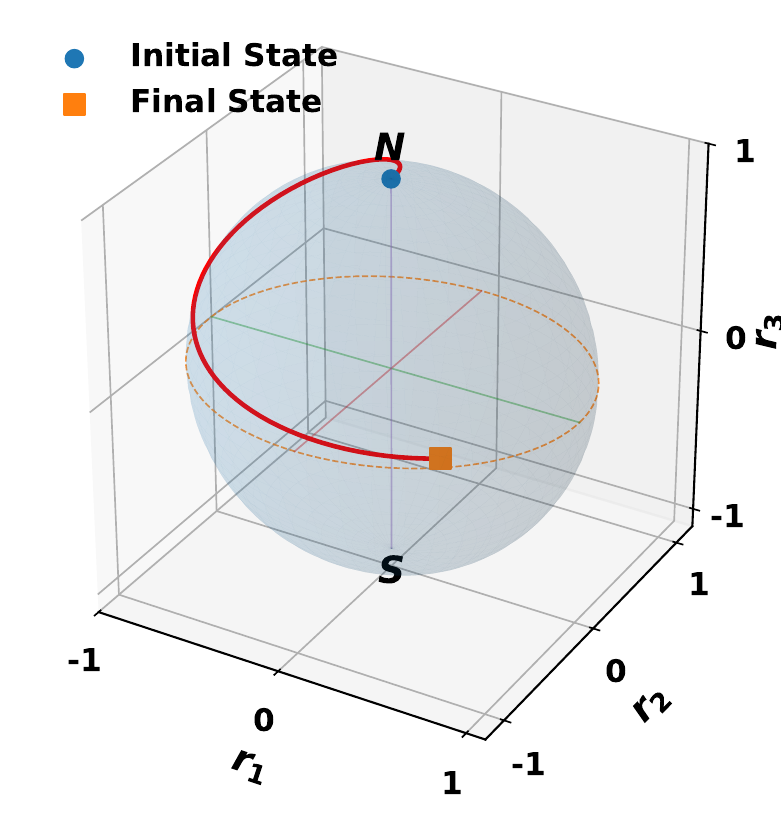}
       \caption{Left panel shows the graph of the potential~\eqref{v=1} as a function of $x$ for $v_0,v_1,\sigma$ given by \eqref{specs1}. $x$ and $v(x)$ are given in units of $\ell$ and $\ell^{-2}$, respectively. Right panel shows the plot of the trajectory of the evolving state $\sR(x)$ in $S^2$ for $\sR(0)=\sH_+$, which corresponds to the north pole marked by $N$, and $k=2\ell^{-1}$.}
        \label{fig1}
        \end{center}}
        \end{figure}

Consider the stationary scattering problem defined by the Schr\"odinger equations~\eqref{sch-eq} for a potential of the form
	\begin{equation}
	v(x):=\left\{\begin{array}{ccc}
	v_0+v_1\,e^{-\frac{(x-\ell/2)^2}{2\sigma^2}}
	&\ \text{for}&x\in[0,\ell],\\
	0&\ \text{for}&x\notin[0,\ell],
	\end{array}\right.
	\label{v=1}
	\end{equation}
where $v_0,v_1$ are real or complex parameters, and $\ell$ and $\sigma$ are real and positive parameters. Comparing \eqref{finite-range} and \eqref{v=1}, we have  $a_-=0$ and $a_+=\ell$. Figure~\ref{fig1} shows the plots of the potential \eqref{v=1} and the trajectory representing the evolution of the initial state $\sR(0)=\sH_+=N$ of the quantum system defined by \eqref{H-def}, \eqref{v=1}, $k=2\ell^{-1}$, and 
	\begin{align}
	&v_0=5\,\ell^{-2},
	&&v_1=5\,\ell^{-2},
	&&\sigma=\frac{\ell}{5}.
	\label{specs1}
	\end{align}
To obtain this trajectory, we solved the corresponding time-dependent Schr\"odinger equation~\eqref{sch-eq-t} subject to the initial condition $\Psi(0)=\Phi_+$ numerically\footnote{The numerical calculations were implemented in Python using NumPy and SciPy, with the time evolution obtained via the $\mbox{solve}_-$ivp routine and the DOP853 integration method.}, found the evolving state vector $\Psi(x)$, and substituted the result for $\Psi$ in \eqref{rho=} to determine $\vec r(x)$.

\begin{figure}
	\begin{center}
\includegraphics[scale=.5]{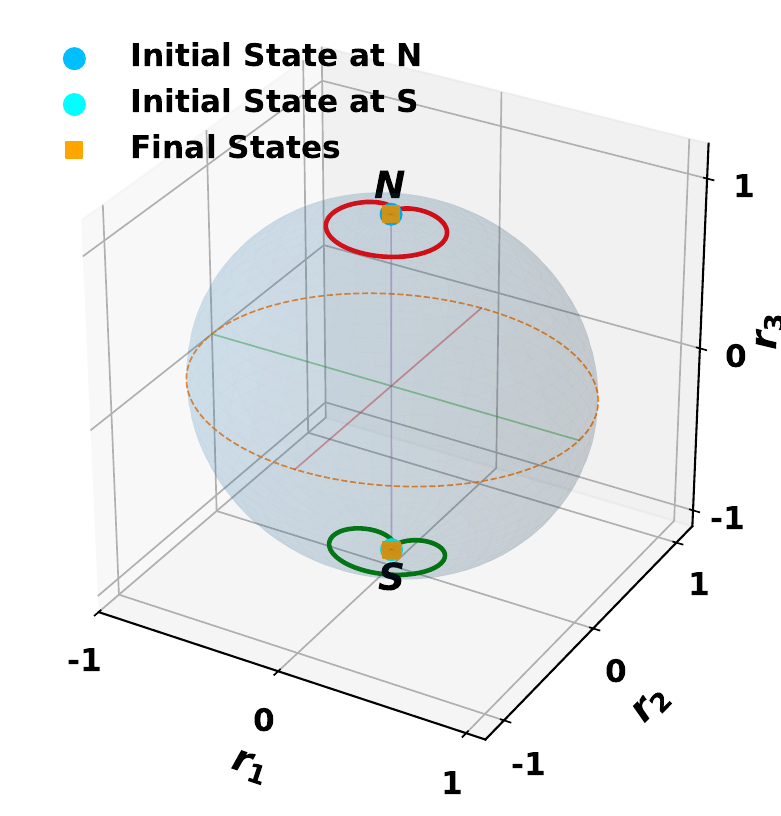}
\caption{Trajectories of the cyclic states of the quantum system corresponding to the potential \eqref{v=1} with initial states $\sR(0)=\sH_+=N$ and $\sR(0)=\sH_-=S$. The parameters $v_0$ and $v_1$ of the potential are given by \eqref{specs1}, and $k=4.88\ell^{-1}$.}
\label{fig2}
\end{center}
\end{figure}

For the above values of the parameters of the potential \eqref{v=1}, the initial state $\sH_-=S$ of the effective quantum system undergoes a cyclic evolution with period $\ell$ for $k=4.88\ell^{-1}$. This shows that $R^r=0$, i.e., the potential is reflectionless from the right at this wavenumber. Since this is a real potential, this implies that it is bidirectionally reflectionless at this wavenumber. Therefore, the initial state $\sH_+$ is also a cyclic state with period $\ell$. Fig.~\ref{fig2} shows the trajectories representing the cyclic evolution of the initial states $\sR(0)=\sH_\pm$ in the Bloch sphere. 

\subsection{Scattering by a piecewise constant complex barrier potential} 

Consider the exactly solvable complex barrier potential 
	\begin{equation}
	v(x):=\left\{\begin{array}{ccc}
	\fz &\ \text{for}&x\in[0,\ell],\\
	0&\ \text{for}&x\notin[0,\ell],
	\end{array}\right.
	\label{v=2}
	\end{equation}
where $\fz$ and $\ell$ are respectively complex and real parameters and $\ell>0$, \cite{tjp-2020,pra-2011a}. This potential describes the scattering of normally incident TE(M) waves of wavenumber $k$ by a homogeneous optical slab of thickness $\ell$ and relative permittivity $\hat\epsilon(x)=1-k^{-2}v(x)$. Positive values of the imaginary part of $\fz$ correspond to situations where the slab is made of a gain material \cite{silfvast} in which case it can satisfy the laser threshold condition \cite{silfvast}. This happens precisely at the spectral singularities of the potential \cite{pra-2011a}. As we described in the preceding section, such a spectral singularity corresponds to a dynamical spin flip that is associated with the evolution of the initial state $\sH_-$ of the effective two-level quantum system given by \eqref{H-def} into $\sH_+$ as $x$ changes from $a_-:=0$ to $a_+:=\ell$. The time-reversed system obtained by changing the sign of the imaginary part of $\fz$, i.e., complex-conjugating the potential, displays CPA, \cite{tjp-2020}. This corresponds to another dynamical spin flip that evolves $\sH_+$ into $\sH_-$. 

\begin{figure}
    	\centering
    	\includegraphics[width=0.4\linewidth]{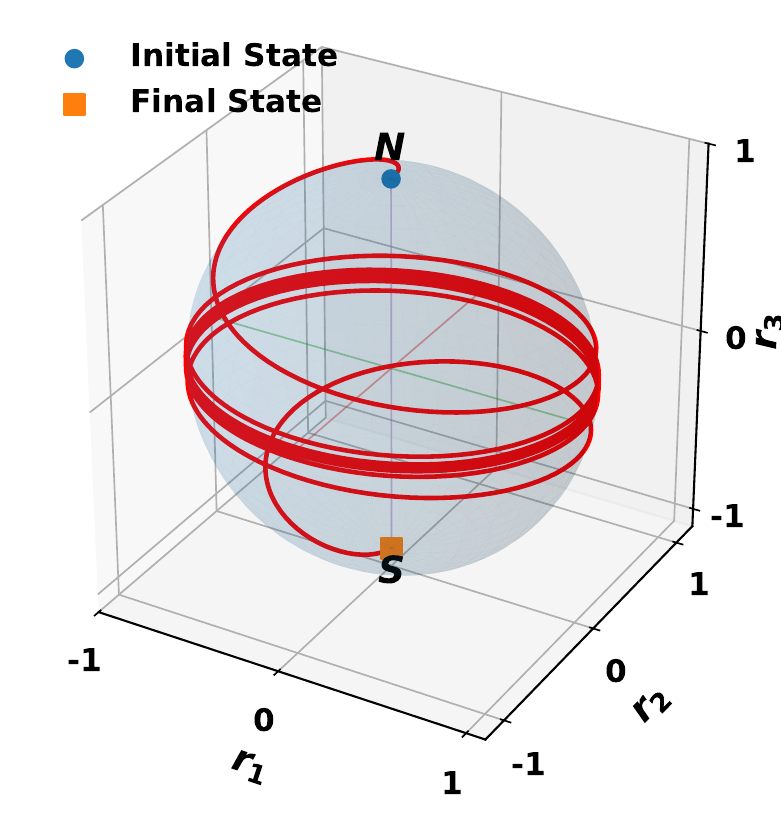}~~~~~
	 \includegraphics[width=0.4\linewidth]{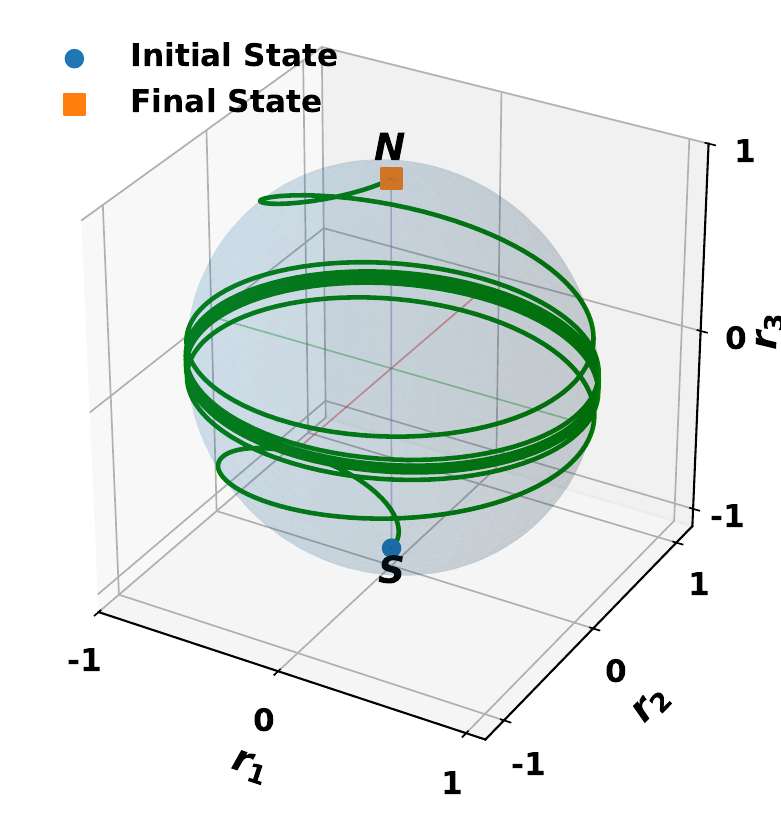}
    \caption{Graphical demonstration of dynamical spin flips associated with the spectral singularity of the potential \eqref{v=2} with $\fz=(617.77-1.56 i)\ell^{-2}$ that occurs at the wavenumber $k=25.05\ell^{-1}$ (on the left), and its time-reversal dual which corresponds to the CPA realized by the potential \eqref{v=2} with $\fz=(617.77+1.56 i)\ell^{-2}$ for the same wavenumber (on the right).}
    \label{fig3}
\end{figure}

Ref.~\cite{pra-2011a} derives the following condition for the emergence of spectral singularities for the potential \eqref{v=2}.
	\be
	e^{-2i\fn\ell k}=\left(\frac{\fn-1}{\fn+1}\right)^2,
	\label{ss-2}
	\ee
where $\fn:=\sqrt{1-\fz/k^2}$. We examined the spectral singularity corresponding to $\fz=(617.77-1.56 i)\ell^{-2}$ and $k=25.05\ell^{-1}$ which satisfy \eqref{ss-2}. Figure~\ref{fig3} shows the plots of the trajectories of the evolving states associated with the spin flips representing this spectral singularity and its time-reversal dual which is obtained for $\fz=(617.77+1.56 i)\ell^{-2}$ and the same wavenumber  \cite{jpa-2012}.

\subsection{Scattering of TE waves by a unidirectionally reflectionless slab}

Consider an effectively one-dimensional nonmagnetic dielectric slab with a relative permittivity of the form
	\be
    	\hat\varepsilon(x)=1+
    	\left\{\begin{array}{ccc}
    	-2\kappa^2 \cos^2\theta_\star f(x)&\for&x\in[0,\ell],\\
    	0&\for&x\notin[0,\ell],
    	\end{array}\right.
	\label{hat-eps=}
    	\ee
where $\kappa$ and $\theta_\star$ are real parameters, $\kappa>0$,  
	\be
	f(x):= \frac{2x(\ell - x) + i(k_\star|\cos \theta_\star|)^{-1}(\ell - 2x)}{[\kappa^2 x(\ell - x) + 1]^2},
    	\label{f=}
	\ee
and $k_\star$ is real and positive parameter. 

\begin{figure}
    \centering
    \includegraphics[width=0.45\linewidth]{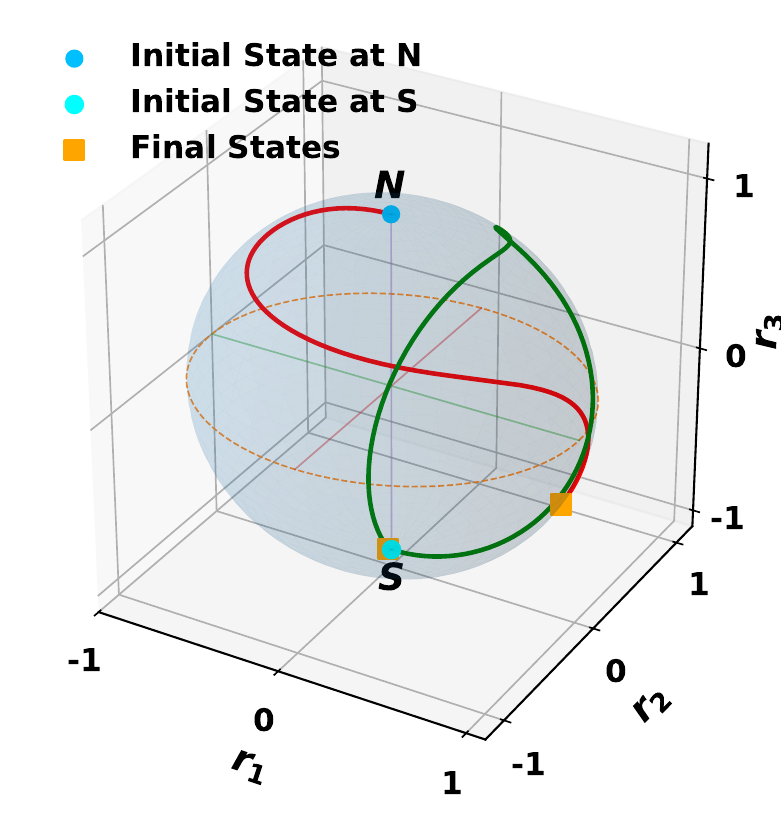}
    \caption{Trajectories associated with the evolution of the initial states $\sH_+$ (in red) and $\sH_-$ (in green) determined by the effective Hamiltonian \eqref{H-def} with $v(x)$ satisfying \eqref{v=3} and $\hat\varepsilon(x)$ and $k$ given by \eqref{hat-eps=}, \eqref{f=}, and \eqref{specsbeta}. The second is a closed curve signifying the cyclic evolution of $\sH_-$ with period $\ell$ while the first does not close as $x$ changes from $0$ to $\ell$. These show that $R^r=0$ and $R^l\neq 0$, respectively.}
    \label{fig4}
\end{figure}

Ref.~\cite{ptep-2024b} shows that scattering of TE waves by such a slab is modeled using the Bergmann equation~\eqref{Bergmann-eq} with $\alpha(x)=1$ and $\beta(x)=\hat\varepsilon$, and that this slab does not reflect oblique TE waves with incidence angle $\theta=\theta_\star\in(90^\circ,270^\circ)$ and wavenumber $k=k_\star$. Because for $\alpha=1$ and $\beta=\hat\varepsilon$, the Bergmann equation \eqref{Bergmann-eq} coincides with the Schr\"odinger equation \eqref{sch-eq} with the potential 
	\be
	v(x)=k^2\cos^2\theta_\star[1-\hat\varepsilon(x)],
	\label{v=3}
	\ee 
this potential is reflectionless from the right for $k=k_\star$, but it turns out not to be reflectionless from the left \cite{ptep-2024b}. In terms of the dynamics generated by the corresponding effective Hamiltonian \eqref{H-def}, this means that the initial state $\sH_-$ undergoes a cyclic evolution with period $\ell$, while $\sH_+$ does not. Figure~\ref{fig4} provides a graphical confirmation of this behavior for 
	\begin{align}
    	&\theta_\star=120^\circ,
	&&\kappa=3\,\ell^{-1},
	&&k=k_\star=4.35\,\ell^{-1}.
    	\label{specsbeta}
	\end{align}

\subsection{Scattering of TM waves by a unidirectionally reflectionless slab}

Consider an effectively one-dimensional nonmagnetic dielectric slab with the relative permittivity 
	\be
    	\hat\varepsilon(x)=1+
    	\left\{\begin{array}{ccc}
    	\fz_0+\fz_1 e^{-\kappa(\ell-x)}&\for&x\in[0,\ell],\\
    	0&\for&x\notin[0,\ell],
    	\end{array}\right.
	\label{alph=}
    	\ee
where $\fz_0$ and $\fz_1$ are complex parameters, and $\kappa$ and $\ell$ are real and positive parameters. As noted in Ref.~\cite{ptep-2024b}, the scattering of TM waves by this slab is modeled using the Bergmann equation \eqref{Bergmann-eq} with 
	\begin{align}
	&\alpha(x)=\hat\varepsilon(x),
	&&\beta(x)=1, 
	\label{alpha-beta}
	\end{align}

\begin{figure}
    \centering
    \includegraphics[width=0.3\linewidth]{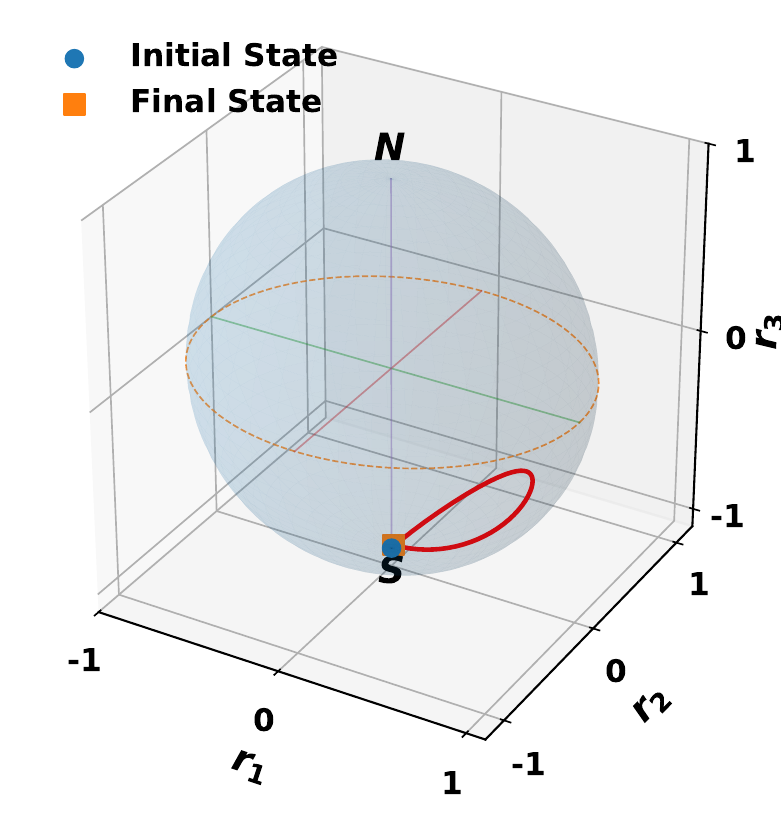}
    ~~\includegraphics[width=0.3\linewidth]{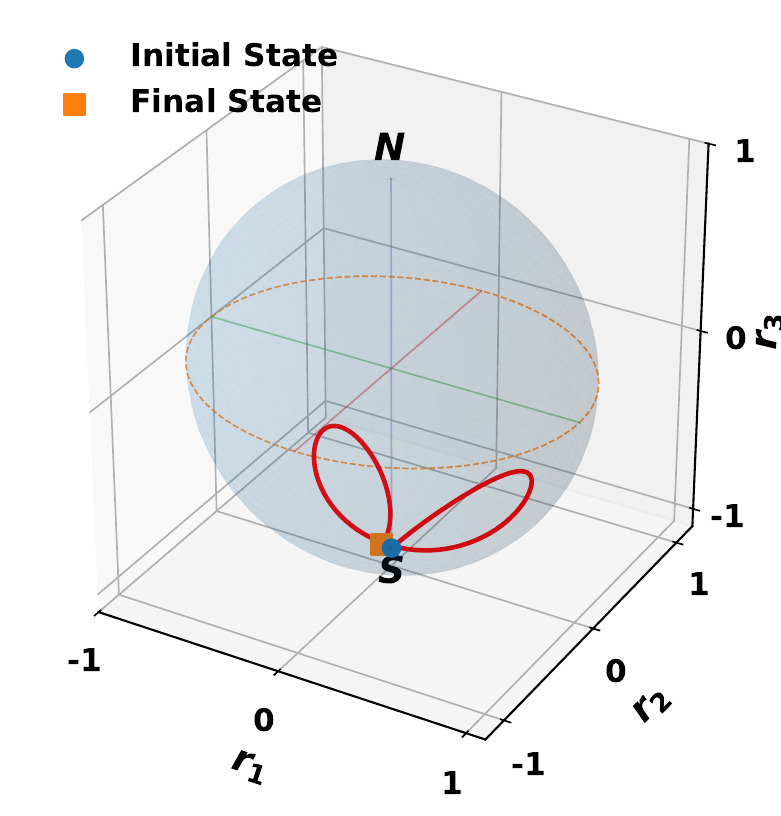}~~
\includegraphics[width=0.3\linewidth]{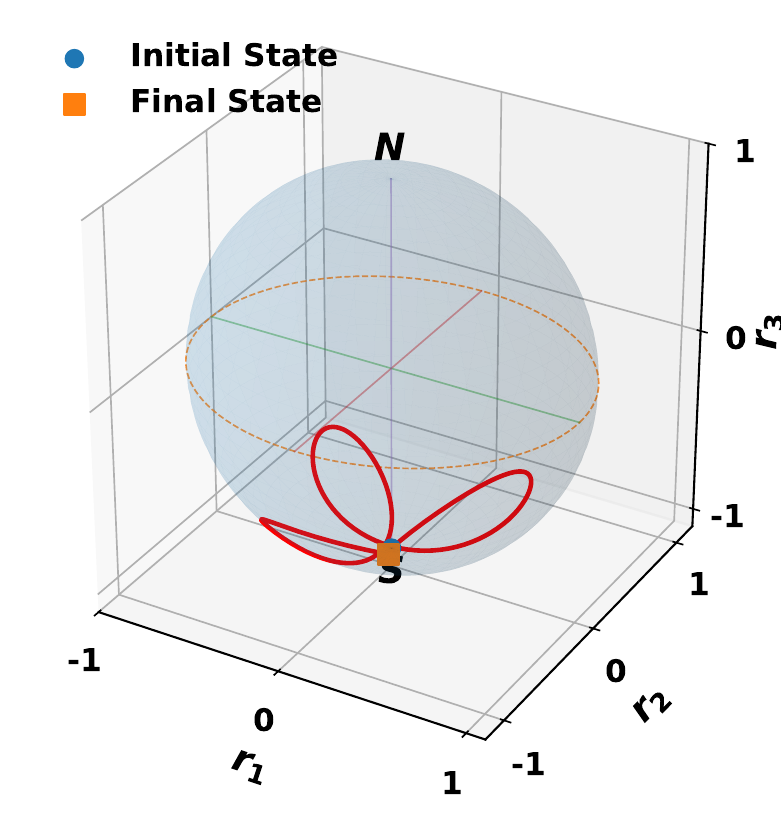}
    \caption{Trajectories of cyclic states of the effective quantum system corresponding to the scattering of TM waves with incidence angle $\theta=120^\circ$ and wavenumbers $k=0.82\ell^{-1}$ (left panel), $k=1.65\ell^{-1}$ (middle panel), and $k=2.48\ell^{-1}$ (right panel) by the permittivity profile given by \eqref{alph=} -- \eqref{specsalpha}.}
    \label{fig5}
\end{figure}

For 
	\begin{align}
	&\fz_0=10+0.1 i,
	&&\fz_1=0.5+0.1 i
	&&\kappa=\ell^{-1},
	\label{specsalpha}
	\end{align}
the permittivity profile \eqref{alph=} turns out to display unidirectional reflectionlessness for TM waves with incidence angle $\theta=120^\circ$ and wavenumbers $k=0.82\ell^{-1},
1.65\ell^{-1},2.48\ell^{-1}$. Figure~\ref{fig5} shows the trajectories obtained by evolving the initial state $\sH_-$ for the corresponding effective system that is defined by \eqref{H-def-2} and \eqref{alph=} -- \eqref{specsalpha}. The fact that these are closed curves is a manifestation of the fact that $R^r=0$. For $k=1.65\ell^{-1}$ and $2.48\ell^{-1}$, the corresponding trajectories of the evolving state respectively pass through the initial state, i.e., the south pole of the Bloch sphere, twice or three times as $x$ ranges over the interval $[0,\ell]$. 

\section{Conclusion}

Quantum systems defined by non-Hermitian Hamiltonian operators have attracted a great deal of attention due to their various physical applications  \cite{review,konotop,zhao,Ashida,bender,Wang-2026}. Most of these applications make use of the spectral properties of these Hamiltonians. In contrast, the application of non-Hermitian Hamiltonians within the dynamical formulation of stationary scattering makes use of the dynamics generated by the associated effective non-Hermitian Hamiltonians.  In the present article, we study the prospects of deriving information about the scattering data by performing quantum measurements on the quantum systems defined by these Hamiltonians. In one dimension, these are two-level system. Therefore the evolution of their states defines trajectories in a Bloch sphere. 

Quantum measurements turn out to provide information about the reflection coefficients $|R^{l/r}|^2$ and the quantity $|T^2-R^lR^r|$ which has the interpretation of the modulus of the determinant of the scattering matrix, if we identify it with \cite{springer-2018,muga}
	\[\bS=\left[\begin{array}{cc}
	T & R^r\\
	R^l & T\end{array}\right].\]
We should however recall that the time-evolution operator for the effective quantum system which determines the transfer matrix of the scattering problem contains the knowledge of $T$ and $R^{l/r}$. This suggests the possibility of gaining information about $T$ and the phases of $R^{l/r}$ by examining the observable quantities associated with the quantum phase factors.  

Among the remarkable outcomes of our investigation is the discovery of the correspondence of the trajectories associated with the emergence of spectral singularities and CPA with the dynamical spin flips that map one of the poles of the Bloch sphere to the other. Another notable observation is that unidirectional reflectionlessness gives rise to situations where one of the poles of the Block sphere undergoes a cyclic evolution while the other traces an open curve.

\subsection*{Acknowledgements} 
This work has been supported by the Scientific and Technological Research Council of T\"urkiye (T\"UB\.{I}TAK) in the framework of the project 123F180 and by Turkish Academy of Sciences (T\"UBA).

\ed